\documentclass{elsart}
\newcommand{\degree}{\ensuremath{^\circ}}

\usepackage[sort&compress]{natbib}
\usepackage{threeparttable}

\usepackage{graphics}
\usepackage{graphicx}
\usepackage{epsfig}

\usepackage{amssymb}

\begin{document}

\begin{frontmatter}



\title{HHSMT Observations of the Venusian Mesospheric Temperature, Winds, and CO abundance around
the MESSENGER Flyby}


\author[label1]{Miriam Rengel\corauthref{cor}} \corauth[cor]{Corresponding
author.}, \ead{rengel@mps.mpg.de}
\ead[url]{http://www.mps.mpg.de/homes/rengel/}
\author[label1]{Paul Hartogh},
\author[label1]{Christopher Jarchow}
\address[label1]{Max-Planck-Institut f\"ur Sonnensystemforschung,
                  Max-Planck-Strasse 2, 37191 Katlenburg-Lindau, Germany}
\begin{abstract}
We present submillimeter observations of $^{12}$CO J=3–-2 and
J=2--1, and $^{13}$CO J = 2–-1 lines of the Venusian mesosphere and
lower thermosphere with the Heinrich Hertz Submillimeter Telescope
(HHSMT) taken around the second MESSENGER flyby of Venus on 5 June
2007. The observations cover a range of Venus solar elongations with
different fractional disk illuminations. Preliminary results like
temperature and CO abundance profiles are presented.

These data are part of a coordinated observational campaign in
support of the ESA Venus Express mission. Furthermore, this study
attempts to contribute to cross-calibrate space– and ground-based
observations, to constrain radiative transfer and retrieval
algorithms for planetary atmospheres, and to a more thorough
understanding of the global patters of circulation of the Venusian
atmosphere.

\end{abstract}


\end{frontmatter}

\section{Introduction}\label{intro}

NASA's MESSENGER spacecraft swung by Venus for a second time on 6
June 2007 at 23:10 UTC on its way to Mercury. ESA's Venus Express,
on the other hand, is orbiting around Venus since 11 April 2006.
Both spacecrafts carried out multi-point observations of the
Venusian atmosphere on June 6 for several hours. Among the
space-based observations, a world-wide Earth-based Venus Observation
campaign from 23 May to 9 June 2007 (and later) was initiated to
remotely observe the Venusian
atmosphere\footnote{http://sci.esa.int/science-e/www/object/index.cfm?fobjectid=41012}.
It contributes to the growing information on Venus's atmospheric
characteristics and complement the space-based data. Because Venus
was close to its maximum eastern elongation during the time-frame of
the ground-based observations, Venus was in a favorable position for
observations of both its day and night sides.

The Venusian atmosphere is conventionally divided into three
regions: the troposphere (below 70 km), the mesosphere (70 - 120
km), and the thermosphere (above 120 km). Studying the Venus'
mesosphere dynamics is of special interest because this region is
characterized by the combination of two different wind regimes (a
retrograde super-rotation and a sub-solar to anti-solar flow
pattern), and affects both the chemical stability and the thermal
structure of the entire atmosphere \citep{c03}. The principal
feature of atmospheric general circulation is the super-rotation
with typical wind velocities of 60--120 m s$^{-1}$. Mesospheric
temperatures and CO mixing ratio experience global variations with
time \citep{cm91}, probably due to gravity wave breaking activity
\citep{l94}. Submillimeter spectral line observations play an
important role in the investigation of the poorly constrained Venus
mesosphere (it is the only technique to provide direct wind
measurements in the mesosphere). Carbon monoxide (CO) is an
important tracer in the atmosphere of Venus. Because its relatively
strong transitions and the pressure-broadened lineshapes, it is the
best measured trace component of the mesosphere \citep{k76,w81}.

This paper reports CO observations performed in June 2007 on the
mesosphere of Venus as a part of the ground-based observing campaign
in support of Venus Express and MESSENGER. We present some examples
of the capabilities of these data by the use of radiative transfer
and retrieval simulations: preliminary results of the absorption
line Doppler wind velocities, and thermal and CO abundance vertical
profiles.

\section{Observations}\label{obs}

CO Venus observations were made with the Heinrich Hertz
Submillimeter Telescope (HHSMT), operated and owned by the Arizona
Radio Observatory (ARO).
The telescope is located at an elevation of 3178 m on Mount Graham,
Arizona, and consists of a 10 m diameter primary with a nutating
secondary. The observations were obtained on 8, 9, 10, 14 and 15
June from 18:30 to 0:30 UT. We used the 345 Superconductor-Insulator
Superconductor (SIS) and the 2mmJT/1.3 mmJT ALMA\footnote{developed
as part of the ALMA project, this system is the first of this kind
to incorporate the latest SIS mixer technology: the image-separating
mixers. Here, the image separating system operates truly separating
image noise and signal. It uses an old 1.3mm and 2mm quasioptical JT
Dewar and cross-grid to separate the two orthogonal linear
polarizations.} receivers, operating respectively at 320--375 and
210--279 GHz to observe the CO J = 2–-1 (at a frequency of 230.538
GHz), $^{12}$CO J = 3–-2 (at 345.79 GHz), and $^{13}$CO J = 2--1 (at
220.398 GHz). The 345\,SIS receiver was used in the single sideband
mode with the signal frequency being placed once in the lower
sideband (LSB) and another time in the upper sideband (USB), and the
2mmJT/1.3mmJT one only in the LSB. Here, the mixer itself is
intrinsically a double sideband (DSB) mixer. The mixer is connected
to the same input port at both USB and LSB, and then a DSB receiver
can be used in two modes (to measure narrow-band signals contained
entirely within one sideband, and to measure broadband (or
continuum) sources whose spectrum covers both sidebands). System
temperatures with the 345 GHz receiver were typically 1500–-2500 and
200–-500\,K with the 2mmJT/1.3mmJT receiver. Seven different
backends were used simultaneously: two 1 MHz Forbes filterbanks
(FFBA and FFBB), two 970 MHz wide Acousto-Optical-Spectrometers
(AOSA and AOSB), two filterbackends (FB2A and FB2B) filterbackends, and one 215
MHz CHIRP Transform spectrometer (CTS, resolution of $\sim$40\,kHz)
\citep{h90,v06}.

Observations were carried out during good atmospheric conditions
(low water vapor), although  on 10, 14, and 15 June it was partially
cloudy. The observing mode was always dual beam switching. Pointing
was checked every 2--3 h. The typical integration time
per individual spectrum was around 4\,min.

The angular diameter of Venus was 23.44$''$  at the beginning and
25.55$''$ at end of our campaign, respectively. The fraction of
illumination for Venus was 49.95 and 45.68\%, as seen by observer.
Fig.\,1 shows a synthetic image of the apparent disk of Venus that
approximates the telescopic view of Venus as seen from the Earth at
8 June and 18:30 UT\footnote{http://aa.usno.navy.mil/}.


The CO J = 2–-1 line was mapped on 8 different beam positions on
Venus disk, $^{12}$CO J = 3–-2 line on eight positions, and $^{13}$CO J
= 2--1 line on one. The later one represents the first detection of
this line on a planetary atmosphere at HHSMT. A summary of the observations carried out is
provided in Table\,1. Fig.\,2 shows the mapping of the beam
positions on Venus disk.

\normalsize

\section{Data Analysis}\label{da}

The measured spectra were reduced with the CLASS software package of
the Grenoble Astrophysics
Group\footnote{http://www.iram.fr/IRAMFR/GILDAS}. A total of 36
spectra of Venus were taken.

The CTS is able to handle the strong continuum background from Venus
due to its higher dynamical range larger than 30 dB. Because the
retrieval of the temperature and CO distribution require clean
spectra, this spectrometer is well suitable for our goals. Fig.\,3
shows an example of the spectra morphology for $^{12}$CO J = 2–-1
line for different backends which we have used.

\section{Observational Results}
\subsection{Qualitative wind measurements}
The only method that provides wind measurements is the analysis of
Doppler shifts of molecular lines. Spectral line differences with
the East and West limb positions yields measurements of projected
doppler velocities relative to the disk center (gives morning and
afternoon zonal winds).
Fig.\,4 shows examples of the spectra of CO J = 2–-1 lines (Obs. no.
19, 25, and 28) at three different beam positions (5, 11, and 13).
In this example the derived wind speed does not exceed 100
m\,s$^{-1}$.
The Venus-HHSMT relative velocity at the time each scan is not
computed here.

\subsection{Thermal structure and CO Distribution}

In order to retrieve the temperature profile and the CO distribution
in the mesosphere, we have applied a retrieval technique described
by C. D. Rodgers as optimal estimation \citep{r76}. We used a
radiative transfer code \citep{j95,j98,h04} which describes the
physics of the radiative transfer through the atmosphere, to
calculate the synthetic spectra which best fit the observed spectra.
An a priori profile to be retrieved is required as initial input for
the optimal estimation technique. Our atmospheric model consisted of
30 layers spanning the 40--120 km interval with a resolution of 2
km. Brightness temperatures were convolved with an assumed Gaussian
beam.  Below we present the retrieved temperature and CO vertical
profiles taken in the center of the Venus disk obtained with our
technique. At the other beam positions the results will be discussed
elsewhere.

Examples (corresponding to Obs. nos. 5, 32, and 35 in Table\,1) of
fits to the $^{12}$CO J = 3–-2, CO J = 2–-1 and $^{13}$CO J = 2--1
lines in terms of temperature vertical profile are displayed in
Figs.\,5, 6, and 7, respectively. The spectrum of Obs. no. 35
presents signatures like periodic ripples in the baseline. Although
the reasons of this anomaly is currently unknown (perhaps they are
standing waves or an instrumental effect), we retrieved its thermal
profile and CO distribution as a pure exercise. The baseline
signatures may cause large retrieval errors, and we are aware that
it requires further analysis.

Fig.\,8 presents a comparison of the Obs.\,5 temperature retrieval
to the profiles from the SPICAV onboard Venus Express \citep{n07},
Pioneer Venus (PV) descent probes \citep{s80}, to the OIR sounding
measurements \citep{st83}, and to the PV night probe \citep{sek82}.
The extensive layer of warm air at altitudes 90--120 km detected by
SPICAV \citep{n07} (interpreted as the result of adiabatic heating
during air subsidence) seems to be also detected in the HHSMT
profile at 90-- to 100\,km altitude, but the HHSMT peak shows a
shorter temperature excess with respect to SPICAV measurements. The
measurements with SPICAV for orbits 102--104 taken at altitude
4$\degree$ S, for orbits 95, 96, and 98 at 39$\degree$ N, and
reported here at 0$\degree$ show the layer of warm air at altitudes
of around 95, 100, and 97\,km. In other words, if the adiabatic
heating is a localized phenomena, the layer seems to move up (in
altitude) with the latitude. Additional data at different latitudes
are required. Furthermore, in other altitudes the HHSMT profile
compares favorably to those returned by the previous measurements.
Similar temperature profiles were also observed \citep{l94,c03}.
Furthermore, it was suggested that a 10-15\,K increasing in the
mesospheric temperatures occur over 1-30 day periods, and much large
variations (20-40\,K) over as yet undetermined timescales
\citep{c03}.

\section{Conclusion}
\begin{itemize}
\item We have carried out several CO mm-wave line observations on different beam positions on Venus disk
during June 2007.
\item From spectra of $^{12}$CO J=2--1 and CO J=3--2 we retrieved well-resolved and accurate vertical profile
of temperature and CO mixing ratio for the June 2007 mesosphere of
Venus.
\item The temperature peak detection reported here at 90-100\,km seems to support
the newly found of the extensive layer of warm air detected by
SPICAV onboard Venus Express.
\end{itemize}

Despite the success of the analysis presented here, some points need
further work. More accurate line-of-sight wind velocities on Venus
will be determined, and gravitationally redshift corrected.
A discussion about Venus circulation for this particular period of
time will be given elsewhere later.

\section*{Acknowledgments}

We thank to the staff of the HHSMT for crucial support while
observing, and to Bertaux J-L. and to Montmessin F. for providing us
the SPICAV data parallel to publication.






\bibliographystyle{elsart-harv}
\bibliography{ws-pro-sample}

Table 1: Observation Parameters
\\
Fig. 1: Synthetic image of Venus that approximates the telescopic
view of Venus as seen from the Earth at 8 June and 18:30 UT. Dotted
lines of longitude and latitude are shown on the surface in black,
every 30 degrees, beginning at 0 degrees longitude and latitude.
\\
Fig. 2: Black points show beam positions where the CO spectra were
mapped on the Venus disk (for a 24$''$ disk diameter). Solid lines
indicate the Venus' equator and central meridian. Dashed circles
indicate the approximate FWHM beam diameter. Left upper, right upper
and left lower panels represent the positions for CO J = 2–-1,
$^{12}$CO J = 3–-2, and $^{13}$CO J= 2--1  lines.
\\
Fig. 3: Example of the spectra morphology for the $^{12}$CO J = 2–-1
line for different backends.  Left center is the spectra taken with
the CTS. Right upper, center, and lower panels show the spectra
taken with AOS, FB2, and FFB backends. An integration time of 30 min
was taken.
\\
Fig. 4: East and west limb CO J=2--1 spectra (dot and short dash
lines, respectively) compared to the disk center spectrum (solid
line). Beam positions corresponds to 11 and 13. An upper limit wind
velocity of 100 m s$^{-1}$ is estimated.
\\
Fig. 5: Upper left panel shows the synthetic spectra solution for
Obs. 5, and lower left panel, the difference between the observed
and fitted spectra. Upper and lower middle panels indicate the
retrieved temperature and CO abundance profiles derived from the
spectrum. The gray lines show the initial profiles, and the
horizontal lines are the error bars. Upper and lower right panels
show the averaging kernels, i.e., the sensitivity of the retrieval.
\\
Fig. 6: Solution for Obs.\,32. See caption Fig.\,5.
\\
Fig. 7: Solution for Obs.\,35. See caption Fig.\,5.
\\
Fig. 8: Temperature profile retrieval (Fig.\,5), solid line,
compared to the profile from the stellar occultations with the SPICAV
onboard Venus Express \citep{n07}, PV descent probes \citep{s80},
from the OIR sounding measurements \citep{st83}, and from the PV
night probe \citep{sek82}. The SPICAV measurements were taken at
latitude 39$\degree$\,N for orbits 95, 96, and 98, and latitude
4$\degree$\,S for orbits 102-104. The Pioneer-Venus derived VIRA
reference profile for latitudes $<$30 are indicated by the squares.
The anomalously warm temperatures returned by the Venera\,10 probe
in 1975 are shown as stars symbols. The absolute uncertainty for the
temperatures derived here is $\pm$15\,K.
\\
\\
\begin{table}
\begin{tiny}
\begin{threeparttable}
\begin{tabular}{crrccccc} \hline \hline
Beam          & d$\alpha$\tnote{a} & d$\delta$\tnote{a} & Obs. No. &  Scan No. &  Line  &  Date     &   Receiver/Sideband\\
\tiny Position      &           &           &          &           &        &  June 2007&                    \\
\hline
1  &   12   &  -2  &  1  &  9--11   &  $^{12}$CO J = 3–-2  &  08  & 345 SIS -- USB\\
2  &   -12  &  -3  &  2  &  12--14  &  $^{12}$CO J = 3–-2  &  08  & 345 SIS -- USB\\
3  &   4    &  8   &  3  &  15--20  &  $^{12}$CO J = 3–-2  &  08  & 345 SIS -- USB\\
4  &   9    &  -5  &  4  &  21--26  &  $^{12}$CO J = 3–-2  &  08  & 345 SIS -- USB\\
5  &   0    &  0   &  5  &  29--30  &  $^{12}$CO J = 3–-2  &  08  & 345 SIS -- USB\\
5  &   0    &  0   &  6  &  36--39  &  $^{13}$CO J = 3–-2  &  08  & 2mm/1.3mm ALMA -- LSB\\
5  &   0    &  0   &  7  &  45      &  $^{12}$CO J = 3–-2  &  09  & 345 SIS -- LSB\\
1  &   12   &  -2  &  8  &  47--51  &  $^{12}$CO J = 3–-2  &  09  & 345 SIS -- LSB\\
2  &   -12  &  -3  &  9  &  52--57  &  $^{12}$CO J = 3–-2  &  09  & 345 SIS -- LSB\\
6  &   9    &  -3  &  10 &  58--63  &  $^{12}$CO J = 3–-2  &  09  & 345 SIS -- LSB\\
7  &   -9   &  3   &  11 &  64--69  &  $^{12}$CO J = 3–-2  &  09  & 345 SIS -- LSB\\
5  &   0    &  0   &  12 &  73--74  &  $^{12}$CO J = 3–-2  &  09  & 345 SIS -- USB\\
5  &   0    &  0   &  13 &  80--81  &  $^{12}$CO J = 3–-2  &  10  & 345 SIS -- USB\\
1  &   12   &  -2  &  14 &  82--93  &  $^{12}$CO J = 3–-2  &  10  & 345 SIS -- USB\\
8  &   -12  &  3   &  15 &  94--105 &  $^{12}$CO J = 3–-2  &  10  & 345 SIS -- USB\\
6  &   9    &  -3  &  16 &  107--118&  $^{12}$CO J = 3–-2  &  10  & 345 SIS -- USB\\
9  &   -9   &  3   &  17 &  119--130&  $^{12}$CO J = 3–-2  &  10  & 345 SIS -- USB\\
5  &   0    &  0   &  18 &  131--133&  $^{12}$CO J = 3–-2  &  10  & 345 SIS -- USB\\
5  &   0    &  0   &  19 &  139--140&  $^{12}$CO J = 2–-1  &  14  & 2mm/1.3mm ALMA -- LSB\\
9  &   14   &  -2  &  20 &  141--146&  $^{12}$CO J = 2–-1  &  14  & 2mm/1.3mm ALMA -- LSB\\
10 &   -14  &  3   &  21 &  147--152&  $^{12}$CO J = 2–-1  &  14  & 2mm/1.3mm ALMA -- LSB\\
9  &   14   &  -2  &  22 &  153--156&  $^{12}$CO J = 2–-1  &  14.  & 2mm/1.3mm ALMA -- LSB\\
5  &   0    &  0   &  23 &  164--165&  $^{12}$CO J = 2–-1  &  14  & 2mm/1.3mm ALMA -- LSB\\
9  &   14   &  -2  &  24 &  166--167&  $^{12}$CO J = 2–-1  &  14  & 2mm/1.3mm ALMA -- LSB\\
11 &   19   &  -2  &  25 &  168--173&  $^{12}$CO J = 2–-1  &  14  & 2mm/1.3mm ALMA -- LSB\\
12 &   -14  &  -19 &  26 &  174--176&  $^{12}$CO J = 2–-1  &  14  & 2mm/1.3mm ALMA -- LSB\\
11 &  19    &  -2  &  27 &  177--182&  $^{12}$CO J = 2–-1  &  14  & 2mm/1.3mm ALMA -- LSB\\
13 &   -19  &  3   &  28 &  183--188&  $^{12}$CO J = 2–-1  &  14  & 2mm/1.3mm ALMA -- LSB\\
10 &   -14  &  3   &  29 &  189--194&  $^{12}$CO J = 2–-1  &  14  & 2mm/1.3mm ALMA -- LSB\\
5  &   0    &  0   &  30 &  195--196&  $^{12}$CO J = 2–-1  &  14  & 2mm/1.3mm ALMA -- LSB\\
5  &   0    &  0   &  31 &  199--202&  $^{13}$CO J = 2–-1  &  14  & 2mm/1.3mm ALMA -- LSB\\
5  &   0    &  0   &  32 &  208--217&  $^{12}$CO J = 2–-1  &  15  & 2mm/1.3mm ALMA -- LSB\\
14 &   16   &  -2  &  33 &  218--227&  $^{12}$CO J = 2–-1  &  15  & 2mm/1.3mm ALMA -- LSB\\
15 &   -16  &  +3  &  34 &  228--237&  $^{12}$CO J = 2–-1  &  15  & 2mm/1.3mm ALMA -- LSB\\
5  &   0    &  0   &  35 &  244--293&  $^{13}$CO J = 2–-1  &  15  & 2mm/1.3mm ALMA -- LSB\\
5  &   0    &  0   &  36 &  295--299&  $^{12}$CO J = 2–-1  &  15  & 2mm/1.3mm ALMA -- LSB\\
\hline
\end{tabular}
\begin{tablenotes}
       \item[a] d$\alpha$ and d$\delta$, right ascension and
declination, are the astronomical coordinates of a point on the
celestial sphere when using the equatorial coordinate system. The
earlier coordinate is the celestial equivalent of terrestrial
longitude, and the later one, to the latitude, projected onto the
celestial sphere.
     \end{tablenotes}
  \end{threeparttable}
\label{op}
\end{tiny}
\end{table}
\normalsize

\begin{figure}
\begin{center}
\psfig{file=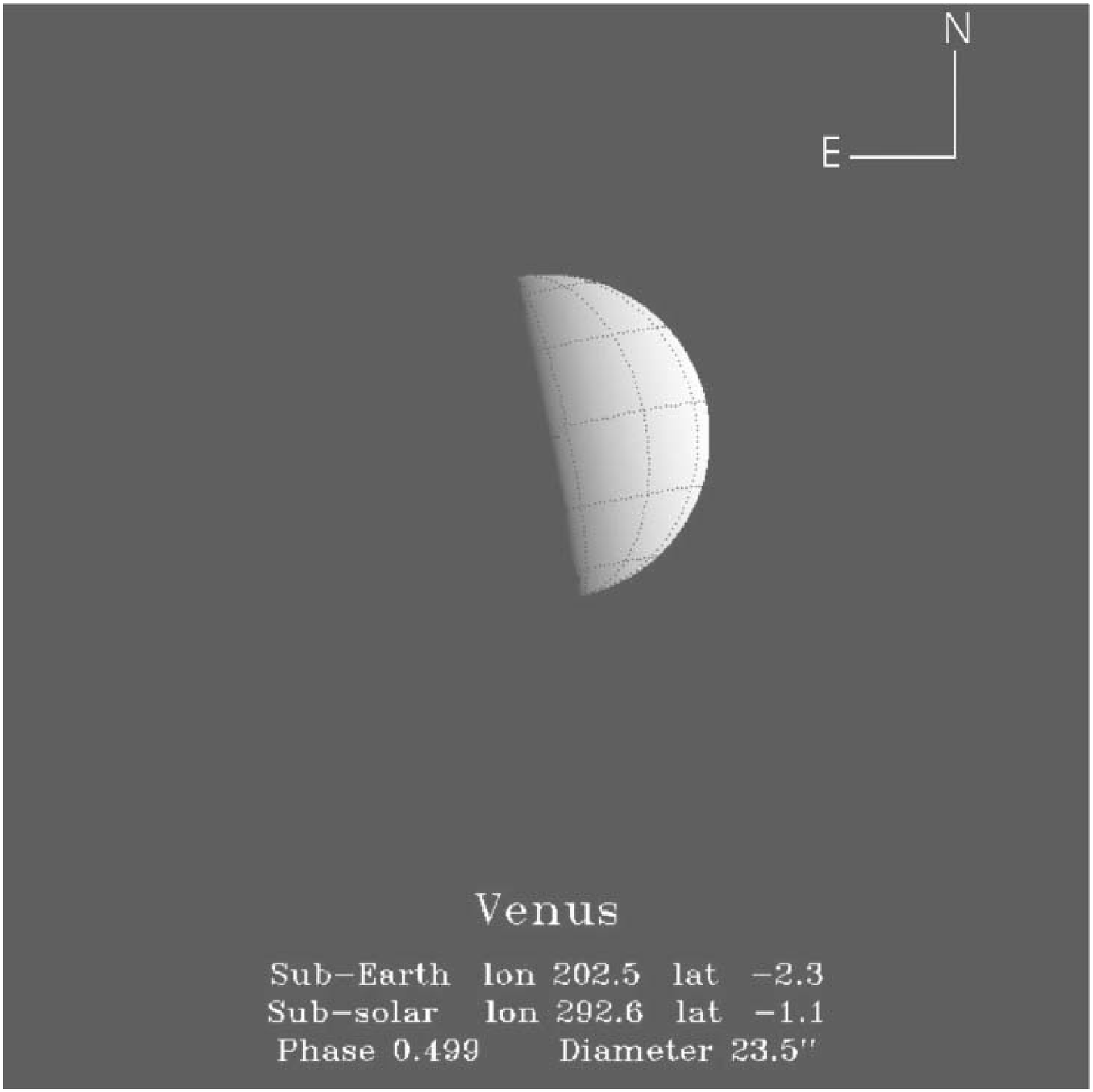,width=6cm}
\end{center}
\caption{} \label{svenus}
\end{figure}

\begin{figure}
\begin{center}
\psfig{file=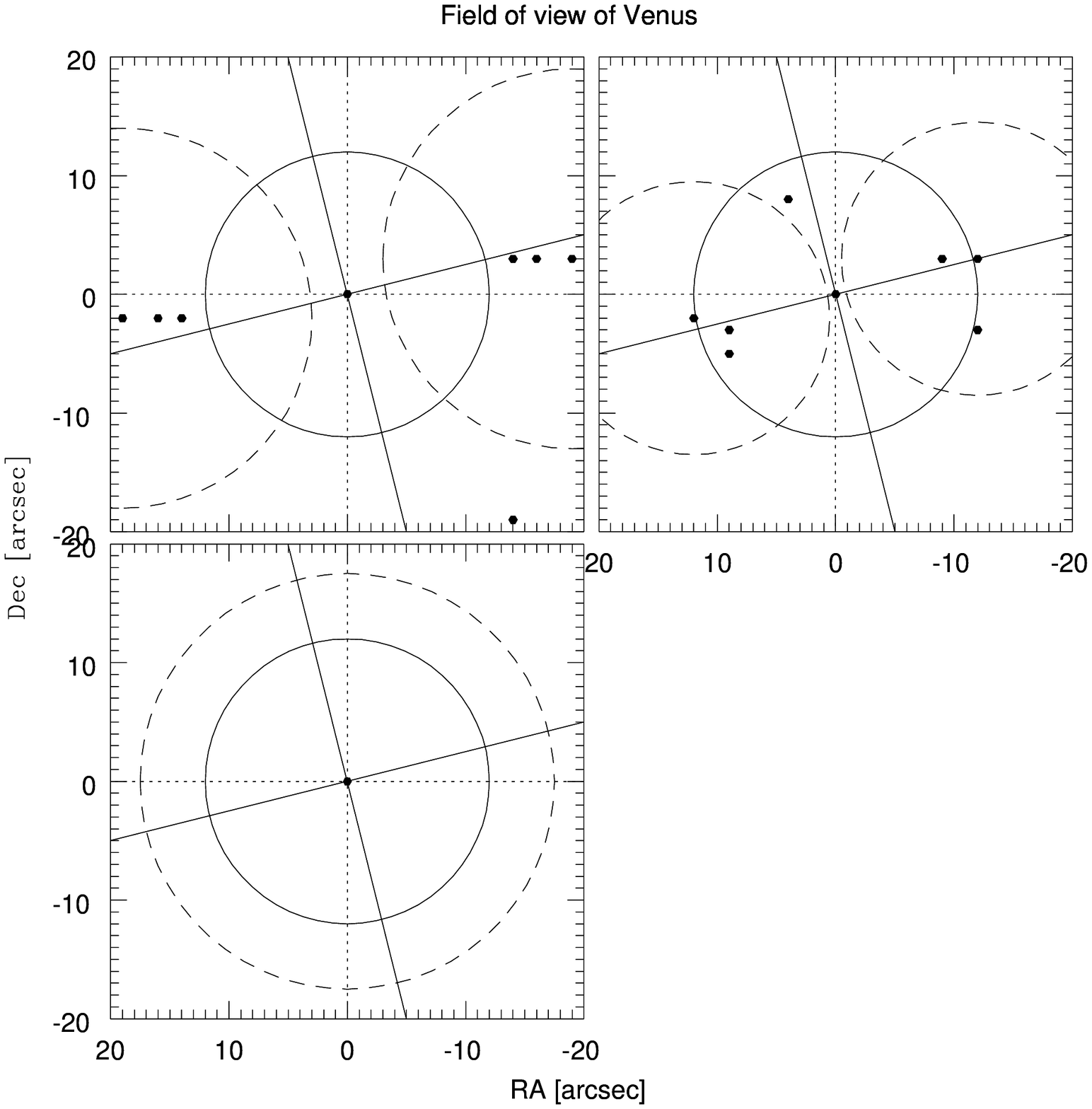,width=7cm}
\end{center}
\caption{}\label{bp}
\end{figure}

\begin{figure}
\begin{center}
\psfig{file=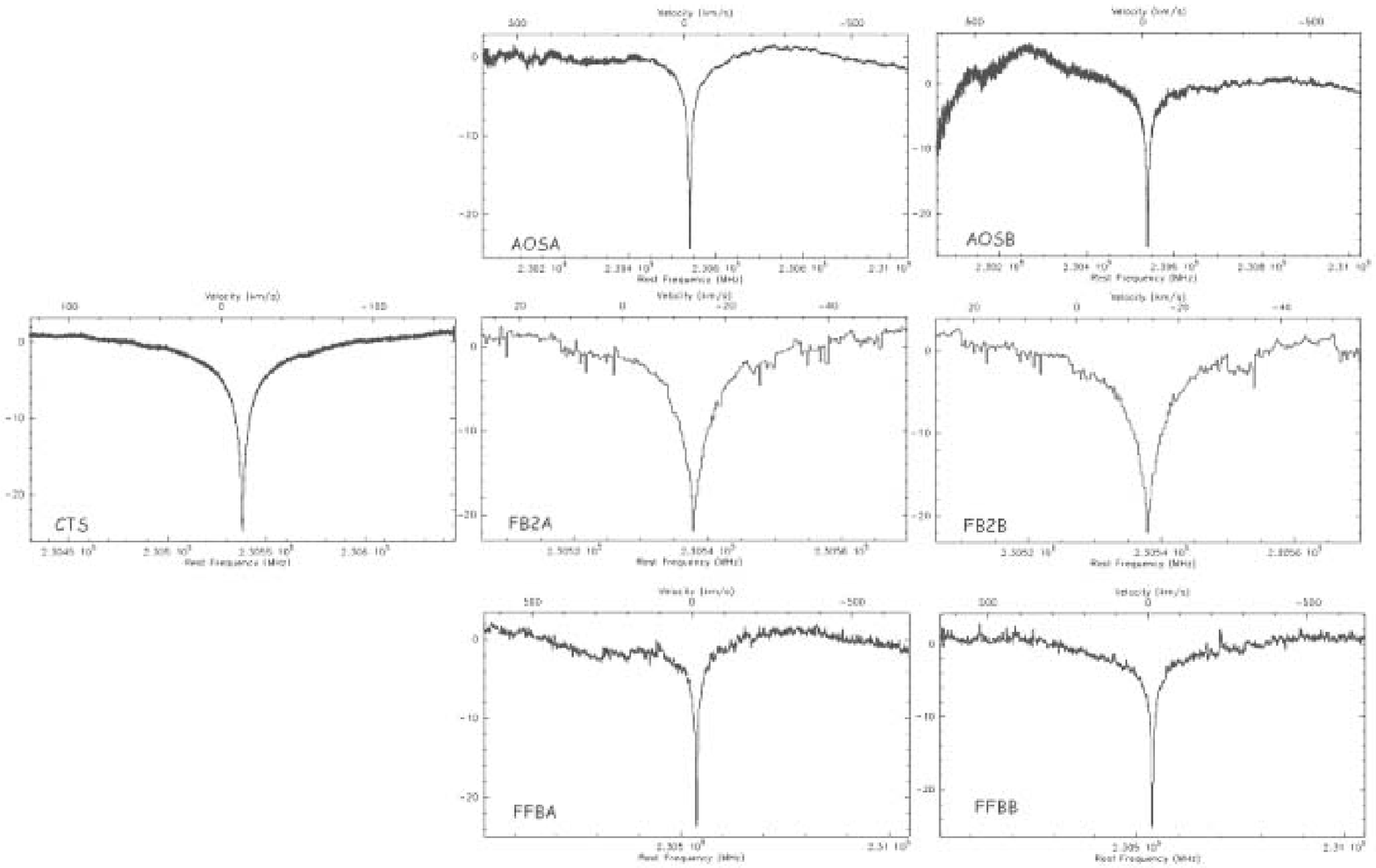,angle=90,width=11cm}\end{center}
\caption{}\label{sp}
\end{figure}

\begin{figure}
\begin{center}
\psfig{file=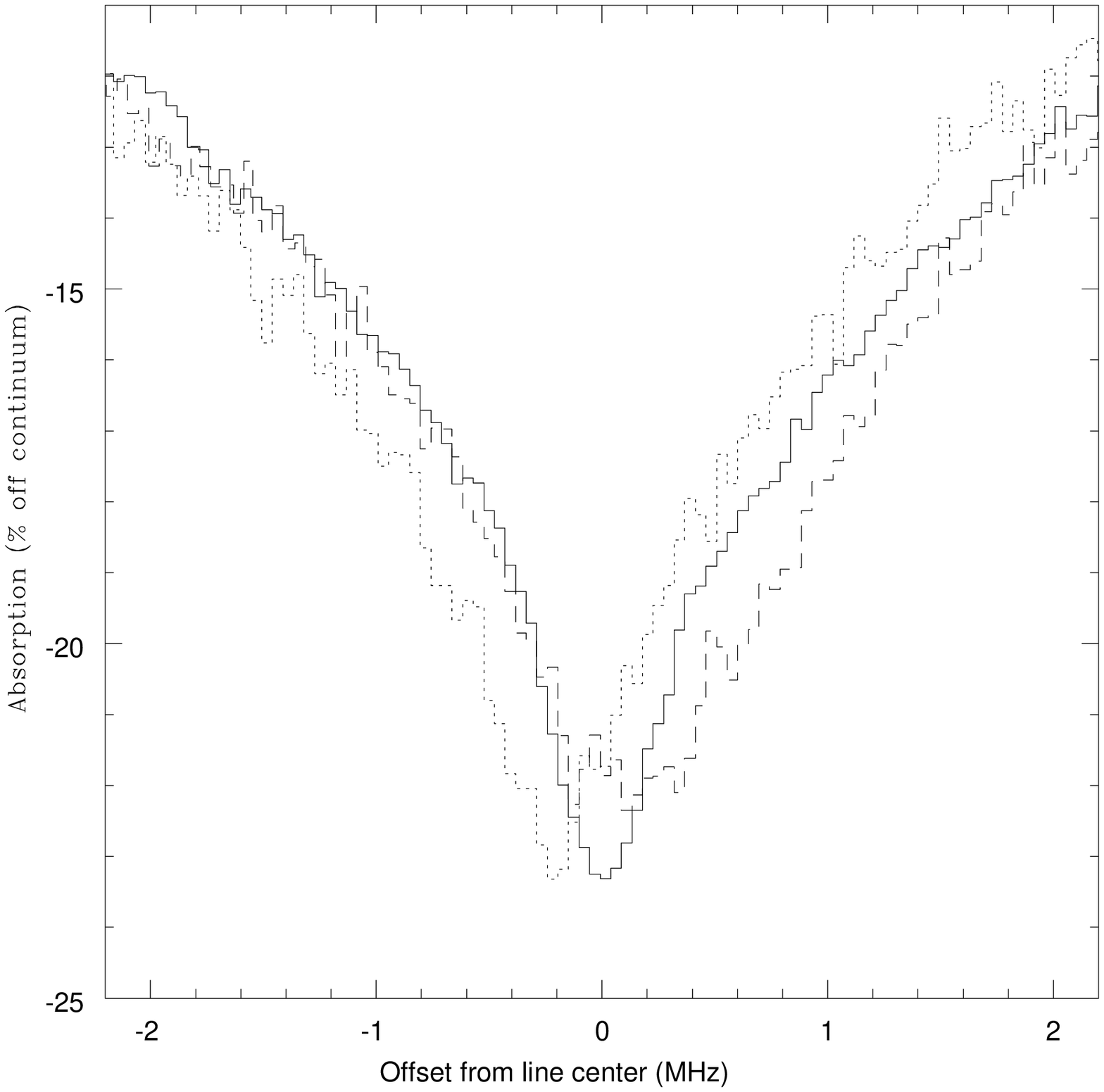,width=6cm}
\end{center}
\caption{} \label{dsg}
\end{figure}

\begin{figure}
\begin{center}
\psfig{file=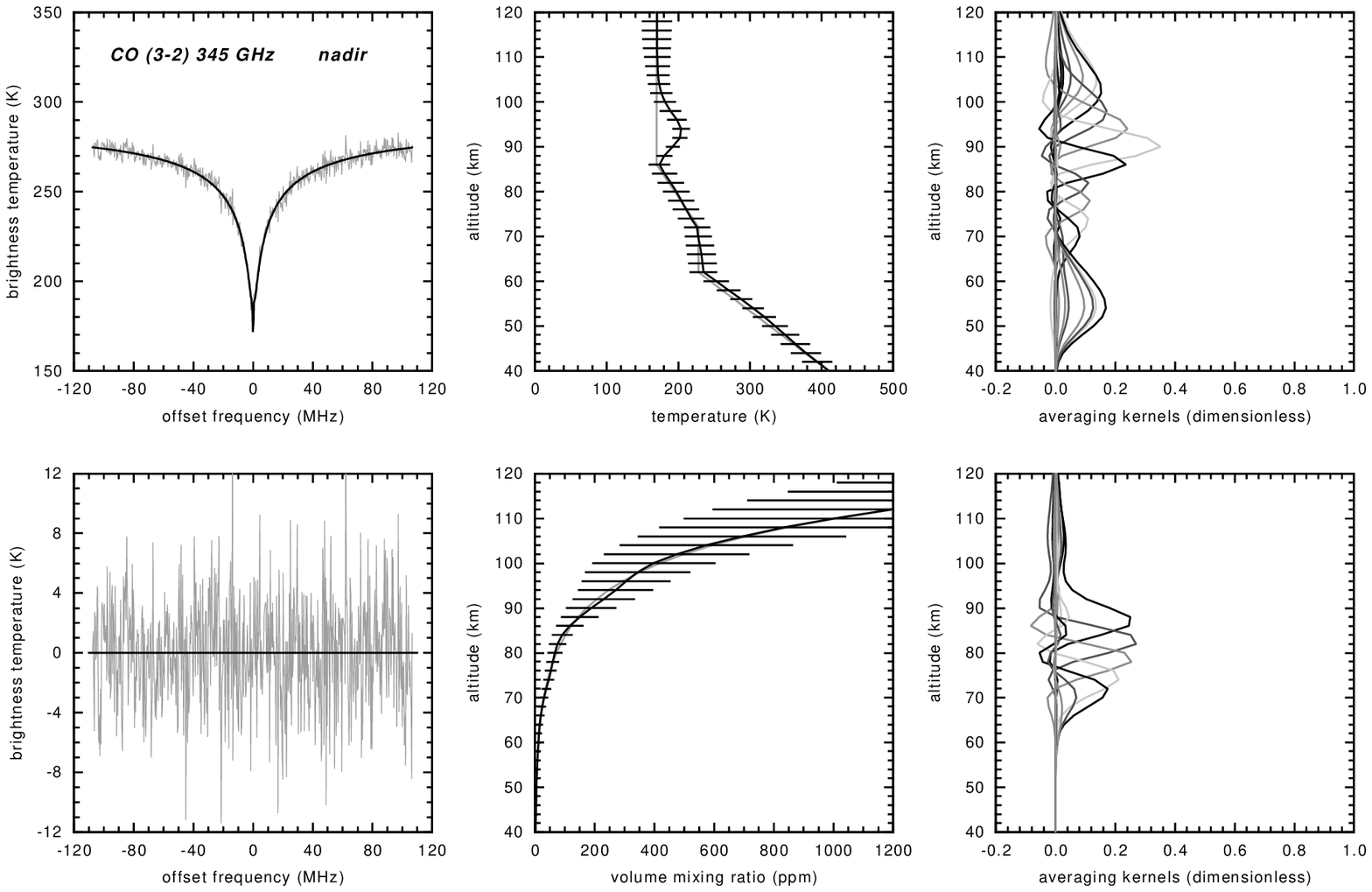,width=10cm}
\end{center}
\caption{} \label{5}
\end{figure}

\begin{figure}
\begin{center}
\psfig{file=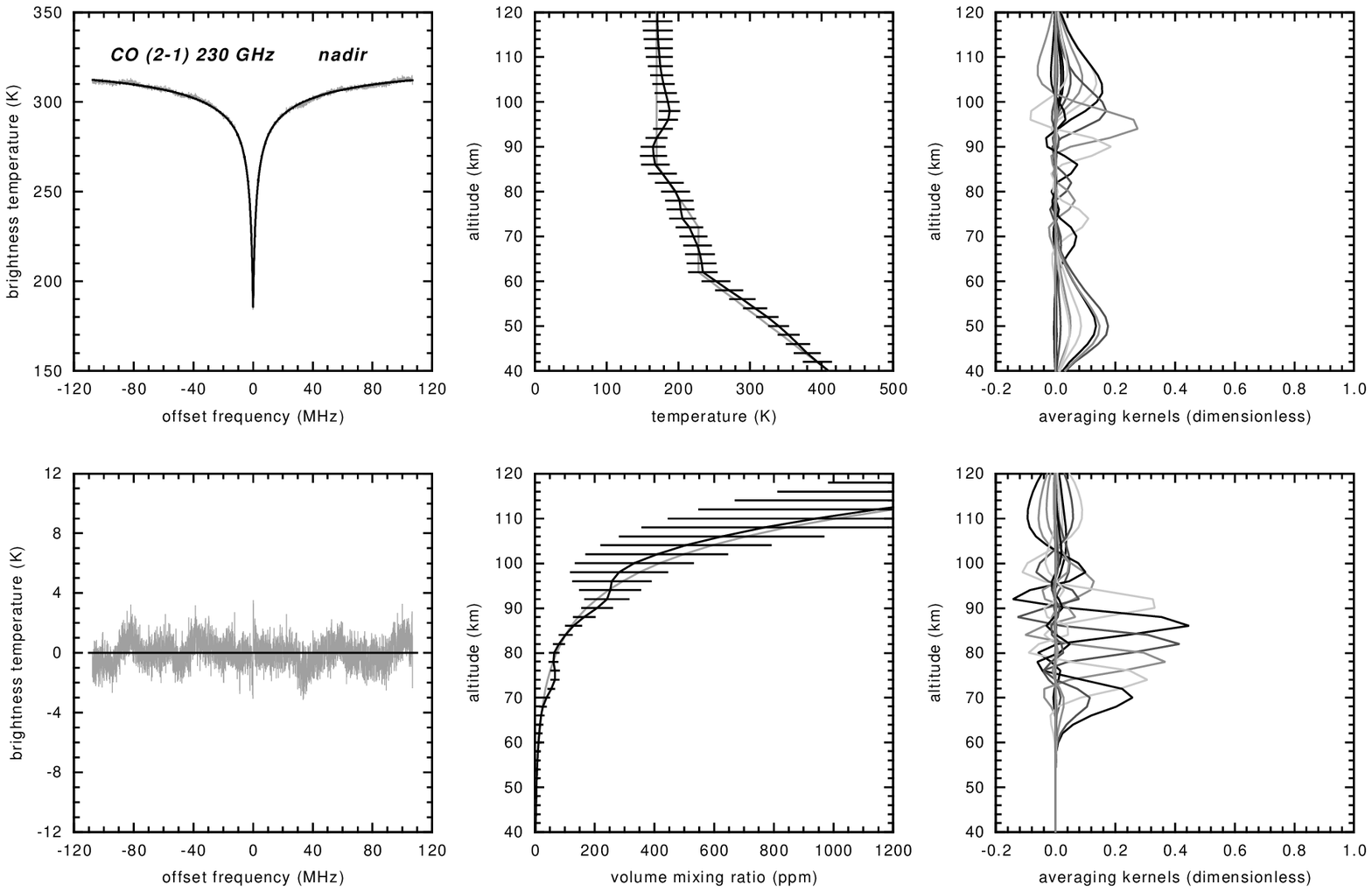,width=10cm}
\end{center}
\caption{}
\label{32}
\end{figure}

\begin{figure}
\begin{center}
\psfig{file=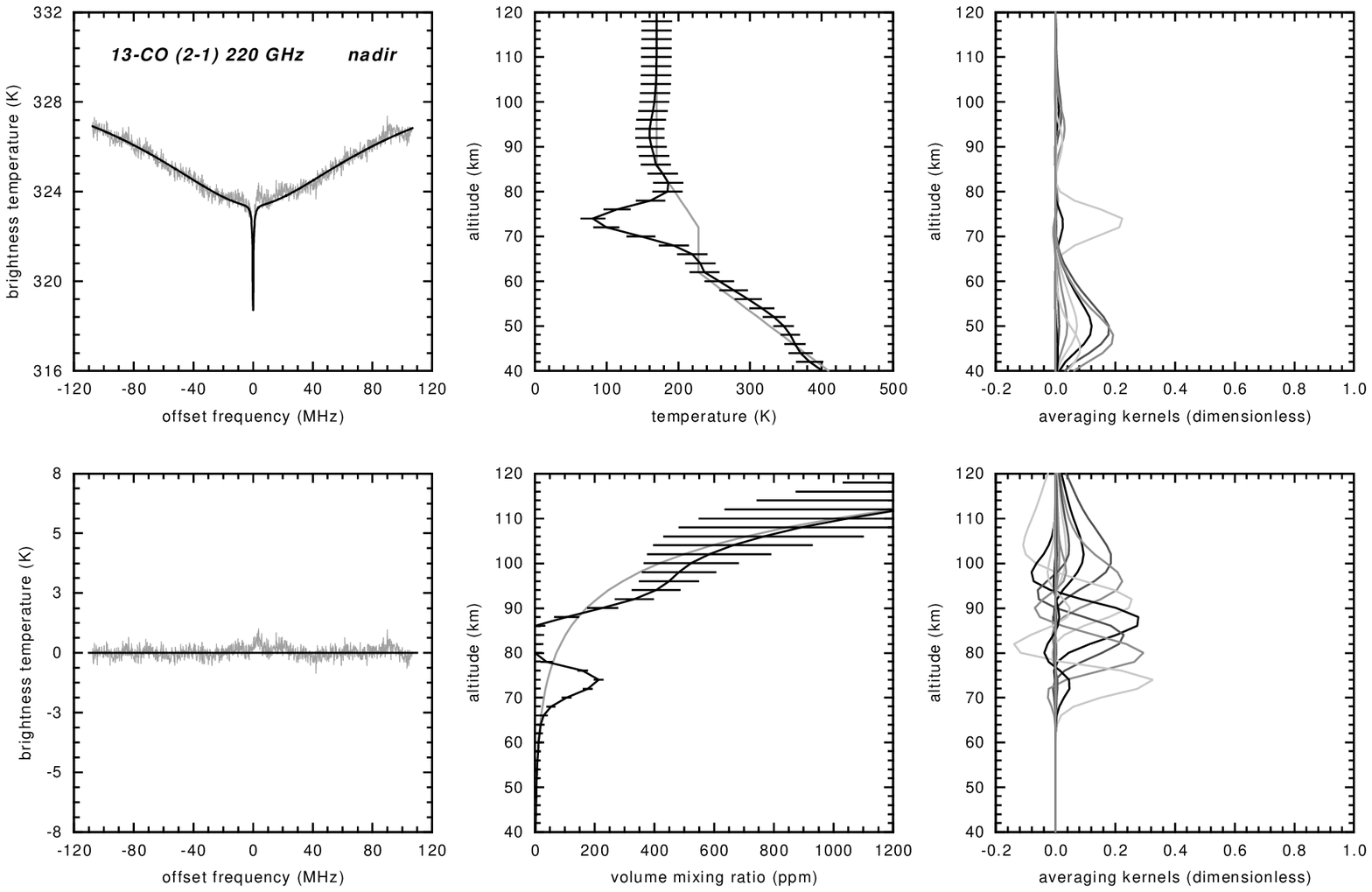,width=10cm}
\end{center}
\caption{}
\label{35a}
\end{figure}

\begin{figure}
\begin{center}
\psfig{file=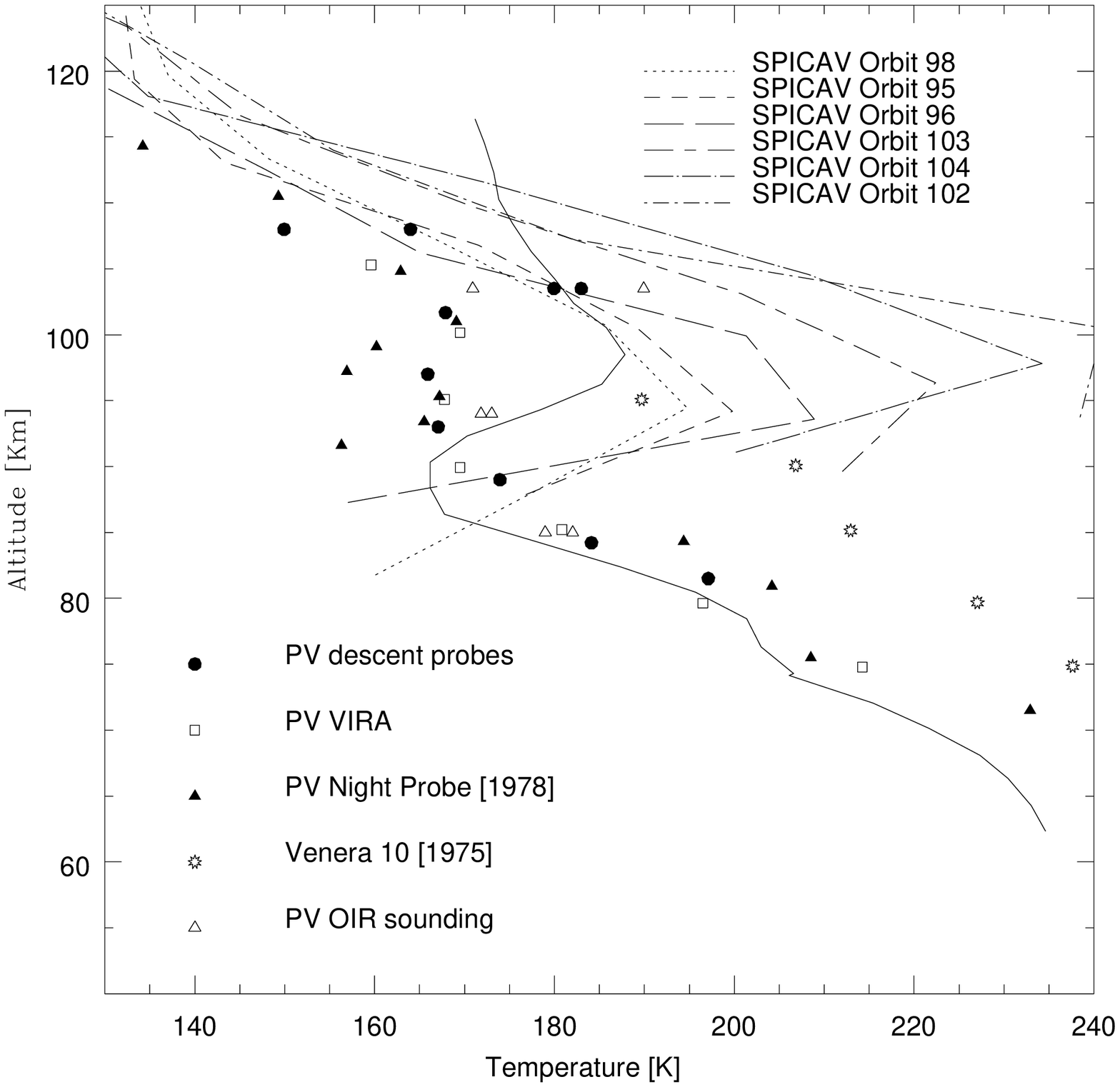,width=8cm}
\end{center}
\caption{}
 \label{compa}
\end{figure}
\end{document}